\documentclass[]{JHEP}
\usepackage{epsfig}
\title{Searching for Narrow Graviton Resonances with the ATLAS
Detector at the Large Hadron Collider}

\author{B.C.~Allanach$^*$, K.~Odagiri$^\dag$,
M.A.~Parker$^\ddag$ and B.R.~Webber$^{\ddag,\$}$\\
$^*$DAMTP, University of Cambridge, Wilberforce Road, Cambridge CB3 0WA,
UK\\
$^\dag$Rutherford Appleton Laboratory, Chilton, Didcot OX11 0QX, UK\\
$^\ddag$Cavendish Laboratory, University of Cambridge, Madingley Road,
Cambridge, CB3 0HE, UK\\
$^\$$Theory Division, CERN, 1211 Geneva 23, Switzerland}

\abstract{
A spectrum of massive graviton states is present in several recent
theoretical models that include extra space dimensions. In some
such models the graviton states are well separated in mass, and
can be detected as resonances in collider experiments. The
ability of the ATLAS detector at the Large Hadron Collider to
identify such states and measure their properties is
considered, in the case that the resonances are narrow compared to
the experimental resolution. The discovery limits for the detection
of the decay mode
$G\rightarrow e^+e^-$ are derived. The
angular distribution of the lepton pair is used to determine the
spin of the intermediate state. In one specific model, the resonance can
be detected up to a graviton resonance mass of 2080 GeV, while the angular
distribution favours a spin-2 hypothesis over a spin-1 hypothesis
at 90\% confidence for resonance masses up to 1720 GeV.
\\
} 


\preprint{Cavendish-HEP-00/07\\
DAMTP-2000-55\\
CERN-TH/2000-158\\
RAL-TR-2000-025\\ hep-ph/0006114}

\begin{document}

\section{Introduction}
The recently proposed localized gravity model of Randall and
Sundrum \cite{randallsundrum} has aroused great theoretical
interest, and many possible extensions and elaborations of this
type of theory are being discussed in the literature~\cite{egs}. 
The model was initially motivated because it solved the weak-Planck scale
hierarchy via an exponentially suppressed warp factor in a
non-factorisable
geometry.
Despite possible problems with negative tension branes and stability in
some of the frameworks, these problems are solved in other models. For
now, we
ignore the details of particular models and concentrate on a 
measurement that should apply to a broad class of the models.
Such
scenarios are characterised by the existence of a
series of graviton excitations, which may be detectable at
present and future colliders. In some cases these excitations may
be sufficiently well spaced to be detected as individual
resonances.

This work will discuss the possibility of
detecting such graviton resonances in as model independent a way as
possible, using the scenario of \cite{randallsundrum} as a guide.
However, the results derived do not depend on the validity of
this particular scenario, but can be applied to any model giving
rise to narrow graviton resonances. 

In the scenario of \cite{randallsundrum} the
massive graviton excitations couple with equal strength to the visible
sector \cite{davoudiasl}. However, the higher modes being suppressed by the
falling parton distribution functions, we consider only the lightest mode. This
does not in any way affect the generality of our approach, as our analysis can
be applied to any such resonances, including the higher modes, so long as the
resonances are narrow and sufficiently separated from the other modes.
This is in contrast to studies in which many excitations, each with small
coupling, contribute to some scattering process~\cite{KKthang}.

An event generator capable of simulating the production and decay
of spin-2 resonances has been developed. This generator is an
extension of the HERWIG 6.1
\cite{HERWIG} simulation package. The generated events are passed
through the ATLAS fast simulation (ATLFAST
\cite{ATLFAST}), in order to give a realistic description of
detector resolution and efficiency. The decay channel
$G\rightarrow e^+e^-$ is chosen for study, since the ATLAS detector
has excellent energy and angular resolution for high energy
electrons, and the Standard Model background comes dominantly
from the well-understood Drell-Yan process. 
The observable cross section for this process is considered first,
and limits on the discovery reach of ATLAS, based simply on
the ability to detect the resonance above the Standard Model
background, are derived. The angular distribution
of the decay products is then considered, in order to study the
ability of the detector to determine the spin of the resonance.

\section{The event generator}
In the model of \cite{randallsundrum}, a 5-dimensional
non-factorizable geometry is used, with two 3-branes of opposite
tension. A graviton Kaluza-Klein spectrum is created, with a scale
$\Lambda_\pi=\bar{M}_{Pl} e^{-kr_c\pi}$, where $\bar{M}_{Pl}$
is the reduced effective 4-D Planck scale, $r_c$ is the
compactification radius of the extra dimension, and $k$ is a
scale of the order of the Planck scale. The
geometrical exponential factor (dubbed the `warp factor')
generates TeV scales from fundamental Planck scales and hence
offers a solution to the hierarchy problem, if $kr_c\approx
12$. The masses of the graviton resonances are given by
$m_n=kx_n e^{-kr_c\pi}=x_n (k/ \bar{M}_{Pl})\Lambda_\pi$ where
$x_n$ are the roots of the Bessel function of order 1. The
couplings of the massive resonances are given by
$1/\Lambda_\pi$. The properties of the model are determined by
the ratio $k/ \bar{M}_{Pl}$. 
We have chosen a value of 0.01 (at the bottom of the range suggested in
\cite{davoudiasl})
for this ratio, which according to \cite{STconstraints}, is on the edge of 95\%
exclusion for a first graviton excitation mass of 
less than 2000 GeV. 
Thus, we assign a low coupling constant to
the gravitons, and hence obtain a conservative estimate of the
production cross section. 
This choice leads naturally to narrow
resonances. This test scenario is used to illustrate the
potential physics reach.

The implementation of the resonance is largely model
independent, depending only on a universal coupling to the Standard Model
fields. This is generally the case for models deriving from extra
dimensions, and the effective Lagrangian is given at first
non-trivial order in
$1/\Lambda_\pi$ as:
 \begin{equation}
  \mathcal{L}_I = -\frac{1}{\Lambda_\pi}h^{\mu\nu}T_{\mu\nu},
 \end{equation}
 where $h^{\mu\nu}$ is the spin-2 field and $T_{\mu\nu}$ is the
energy-momentum tensor of the Standard Model fields.
Here, we consider only the production and measurement of the lightest
massive
graviton excitation. We note however~\cite{randallsundrum}, that other
heavier resonances
are expected. In the case we consider here, they are well split in mass
and
hence the lightest resonance will be produced dominantly.

 The decay $G\to e^+e^-$ is treated in the HERWIG implementation as a
$2\to2$ process, consisting of the two hard production subprocesses
$q\bar q\to e^+e^-$ and $gg\to e^+e^-$. For the $q\bar q\to
e^+e^-$ channel we do not consider interference with the Standard
Model Drell-Yan subprocesses as the interference is negligible for
our purposes. The code is available from the authors on request.

 For calculating the matrix elements and the total decay width of the
graviton resonance, we made use of the results in \cite{han,giudice,hewett}.
We carried out our
calculations both by hand and by using HELAS
\cite{helas} incorporating subprograms with spin-2 particles
\cite{rainwater}. Our implementation of the decay widths corrects
a typographical error in \cite{han} for the decay into heavy
vector bosons but this has only a small effect on the total decay
width.

\begin{equation}
\Gamma(G\to VV)=\delta\frac{m_G^3}{40\pi\Lambda_\pi^2}(1-4r_V)^{1/2}
\left(\frac{13}{12}+\frac{14}{3}r_V+4r_V^2\right),
\end{equation}
where, as in [6], $\delta=1/2$ for identical particles and
$r_V=m_V^2/m_G^2$.

 The angular distributions of the possible subprocesses, in the centre-of-mass
frame of the resonance, are shown in Table \ref{grav_table}. Here $\theta^*$ is the
angle between the decay electron and the beam direction in the dilepton
centre-of-mass frame. The spin-2
distributions contrast strongly with the angular distributions resulting from
spin-1 or scalar resonances. These distributions are shown in Figure
\ref{ang_theory}.

\TABLE{
\caption{Angular distributions in graviton ($G$), vector ($V$) and scalar
($S$) boson production and decay. $\alpha=1$ in Standard Model processes.}
\label{grav_table}
\begin{tabular}{|c|c|}
\hline
Process & Distribution \\
\hline
$gg\rightarrow G\rightarrow e^+e^-$ & $1-\cos^4\theta^*$ \\
$q\bar q\rightarrow G\rightarrow e^+e^-$ & $1-3\cos^2\theta^*
+4\cos^4\theta^*$
\\
$q\bar q$, $gg\rightarrow V\rightarrow e^+e^-$ & $1+\alpha\cos^2\theta^*$ \\
$q\bar q$, $gg\rightarrow S\rightarrow e^+e^-$ & $1$ \\
\hline
\end{tabular}
}

\FIGURE{
%\begin{figure}
%\begin{center}
\hbox{\epsfysize=10cm
\epsffile{ang_theory.epsi}
}
%\end{center}
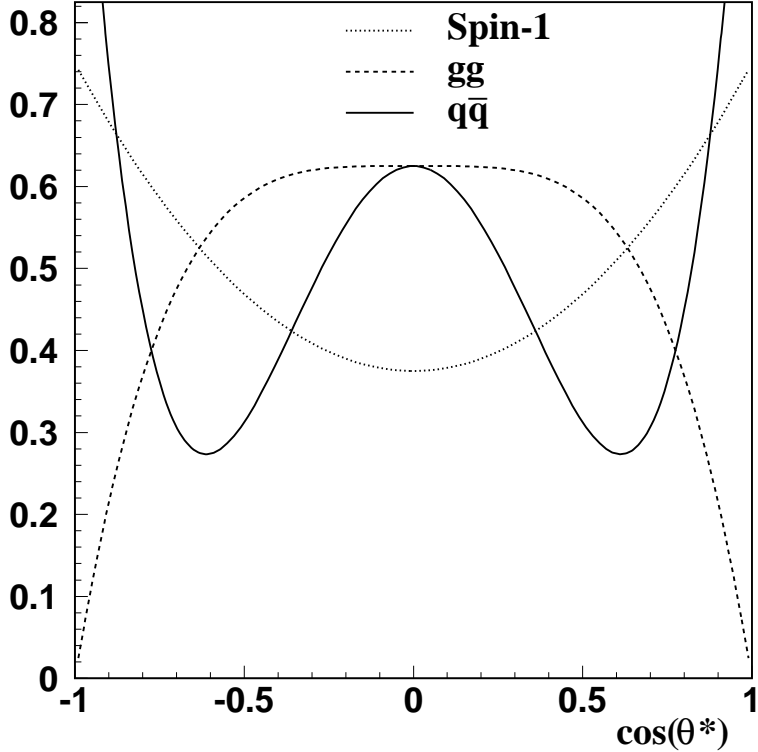
\caption{The normalised theoretical angular distributions for
$gg\rightarrow G \rightarrow e^+e^-$ (dashed curve), $q\bar q\rightarrow
G\rightarrow e^+e^-$ (solid curve) and $q\bar q\rightarrow
Z/\gamma^*\rightarrow e^+e^-$ (dotted curve).}
\label{ang_theory}
%\end{figure}
}

\section{Detection of the graviton resonance}
The process $pp\rightarrow G\rightarrow e^+e^-$ must be
detected above the background from the Standard Model processes
$pp \rightarrow \mathrm{Z}/\gamma^*\rightarrow e^+e^-$. Samples of
events were generated for a single graviton excitation with a mass
ranging from 500 GeV to 2.2 TeV. The masses and widths of the
graviton resonances are given in Table \ref{massres}, as they are computed
with the model parameters given above. The table also shows the
cross-section times branching ratio $\sigma\cdot B$ in the test model and the
width of the resonance reconstructed after the ATLAS detector
simulation. 
%The calorimeter energy resolution is parameterised by
%$\Delta(E)/E = 12\%/\sqrt{E} \oplus 0.245/E_T \oplus 0.007$  where
%E is the electron energy. 
In all the cases studied, the true width
of the resonance is much smaller than the Gaussian experimental
resolution $\Gamma_m$ and can be neglected. Hence the results are
valid for any model in which this condition is met.
In our test model, this is the case for $k/ \bar{M}_{Pl}<0.06$, at
which point the width is approximately equal to the experimental
resolution, and the cross section is 36 times larger than we have
assumed. For larger values of $k/ \bar{M}_{Pl}$, the very high
signal cross sections would overwhelm the Standard Model background,
making the resonance very easy to detect, and the spin simple to
determine, even with a large width.

\TABLE{
\caption{Masses $m_G$ and widths $\Gamma_G$ of the simulated
graviton resonances, and the Gaussian width of the observed
resonance
$\Gamma_m$ after detector effects. $\sigma\cdot B$ is the cross-section
times branching ratio in the test model.}
\label{massres}
\begin{tabular}{llll} \\ 
\hline
$m_G$ (GeV) & $\Gamma_G$ (GeV) & $\Gamma_m$ (GeV) & $\sigma\cdot B$ (fb) \\
\hline
500  & 0.068  & 3.53 & 281.9   \\
1000 & 0.141  & 6.02 & 11.0    \\
1500 & 0.213  & 8.13 & 1.20   \\
1700 & 0.242  & 8.78 & 0.57   \\
1800 & 0.256  & 9.34 & 0.41   \\
1900 & 0.270  & 9.66 & 0.29   \\
2000 & 0.285  & 9.80 & 0.21   \\
2100 & 0.298  &10.18 & 0.15       \\
2200 & 0.312  &10.49 & 0.11       \\
\hline
\end{tabular}
}

For each mass point, having determined the observed resonance
width $\Gamma_m$, a mass window is defined as $\pm 3\Gamma_m$
around the graviton mass. The Drell-Yan cross-section inside this
mass window is then calculated using the HERWIG 6.1 event
generator
\cite{HERWIG}. The number of background events expected in the
mass window is then computed, for an integrated luminosity of 100
fb$^{-1}$, corresponding to one year of LHC running at design
luminosity. The ATLFAST simulation of the ATLAS detector is used
to compute the acceptance for the process, which is dominated by
the pseudorapidity ($\eta$) coverage of the tracker, $|\eta| <2.5$.
Cuts are also made on the isolation of the electrons to account
for the effect of nearby hadrons on the detection efficiency. In
addition to the cuts imposed by ATLFAST, a tracking efficiency of
90\% was applied to each electron, to allow for track
reconstruction losses which are not included in ATLFAST. These losses are
estimated in \cite{ATLAS_TDR} to be approximately 5\% at high transverse
momentum if the tracks are well isolated. Our efficiency value is therefore
conservative. The number of Drell-Yan background events ($N_B$)
inside each mass window after these cuts is given in Table
\ref{DYback}.

\TABLE{
\caption{The number of
signal events in the test model, $N_S$, expected inside each graviton mass
window in a run of 100~fb$^{-1}$, after detector effects; the number of
Standard Model background events, $N_B$; the minimum number of signal
events required to detect the resonance, $N_S^{min}$; and the
minimum production cross-section times branching ratio
$(\sigma\cdot~B)^{min}$ required to detect the
resonance via e$^+$e$^-$ production.}
\label{DYback}
\begin{tabular}{rcrrrc} \\ 
\hline
$m_G$ & Mass window & $N_S$ & $N_B$ & $N_S^{min}$ &
$(\sigma\cdot B)^\mathrm{min}$ \\ 
(GeV)&     (GeV)   &     &      & (fb)\\ 
\hline \\
 500 & $\pm 10.46$ &20750 & 816 & 142.9	& 1.941   \\ 
1000 & $\pm 18.21$ & 814  & 65	 & 40.2	 & 0.542   \\
1500 & $\pm 24.37$ & 84.3 & 11	 & 16.5	 & 0.235   \\
1700 & $\pm 26.53$ & 38.6 & 5.8 & 12.0	 & 0.178   \\
1800 & $\pm 27.42$ & 27.2 & 4.3 & 10.4	 & 0.156   \\
1900 & $\pm 28.29$ & 19.2 & 3.2	& 10.0	 & 0.152   \\
2000 & $\pm 28.76$ & 13.2 & 2.3	& 10.0	 & 0.157  \\
2100 & $\pm 30.55$ & 9.4  & 1.8	& 10.0	 & 0.159  \\
2200 & $\pm 31.46$ & 6.8  & 1.4	& 10.0	 & 0.162   \\
\hline
\end{tabular}
}

The expected signal in the test model for $m_G=1.5$ TeV can be seen in 
Figure~\ref{resonance}, showing the data expected for 100~fb$^{-1}$ of
integrated luminosity. The signal is prominent above the
small Standard Model background. 

\FIGURE{
\hbox{\epsfysize=10cm
\epsffile{resonance1500.epsi}
}
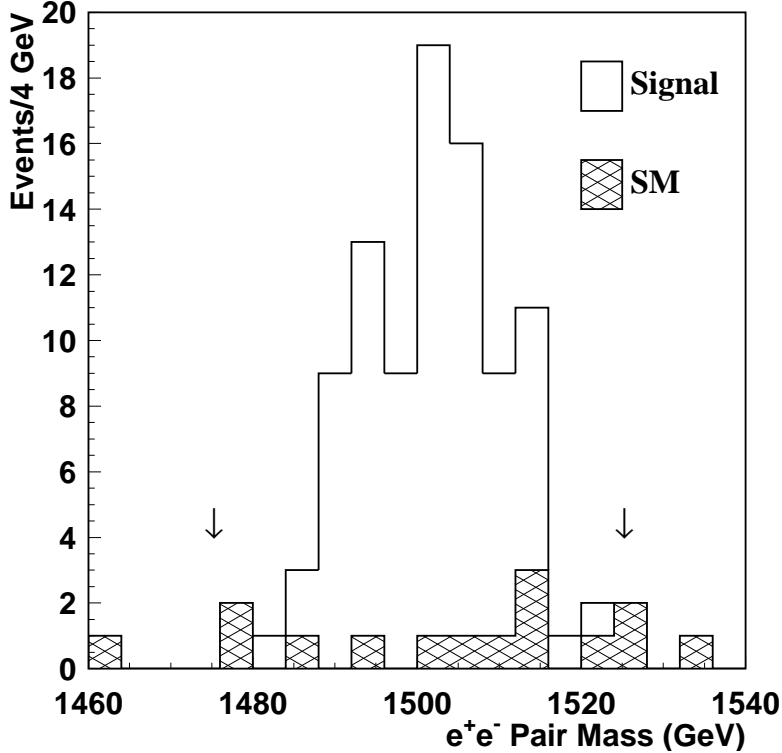
\caption{The number of events per 4 GeV mass bin from a graviton resonance,
with
$m_G=1.5$ TeV (signal), superimposed on the expected Standard Model background
(SM), for 100 fb$^{-1}$ of integrated luminosity. The mass window used to
select the signal is indicated by arrows.}
\label{resonance}
}

We now consider the model-independent detection limit for such a
resonance. The minimum number of signal events, $N_S^{min}$,
needed to detect the resonance above the background is taken to be
$5\sqrt{N}_B$ or 10 events, whichever is
greater, in order to ensure that the signal is statistically
significant. The minimum cross-section times branching ratio
required to produce this number of events is also given in Table
\ref{DYback}, after correcting for the detector acceptance and
efficiency, and assuming an integrated luminosity of 100 fb$^{-1}$.

It can be seen from these results that the background becomes
negligible for graviton masses above 1800 GeV. It should be
emphasised that these results are independent of the test model,
and rely only on the assumption that the width of the resonance to
be detected is negligible compared to the detector mass resolution.
However, as an example, we can compare this cross-section limit
with the values predicted by our test model, in order to determine
the physics reach for ATLAS for this specific case. This is done in
Figure \ref{sigma}, which displays $\sigma\cdot B$ for
the test model and $(\sigma\cdot B)^{min}$. The test
model could be detected for graviton masses up to 2080 GeV. Using
Figure \ref{sigma}, similar estimates of the ATLAS search reach for other
models can be determined without repeating the detector simulation or
background analysis. The search reach for other collider experiments, such as 
CMS, would require a new simulation with the appropriate mass resolution.

Our results differ from those in \cite{ATLAS_TDR} for the case of a $Z^\prime$
with Standard Model couplings, both because of the acceptance difference
arising from the spin of the resonance, and the $Z^\prime$ width, which exceeds
50 GeV for a $Z^\prime$ mass of 2 TeV.

\FIGURE{
\hbox{\epsfysize=10cm
\epsffile{sigma_figure.epsi}
}
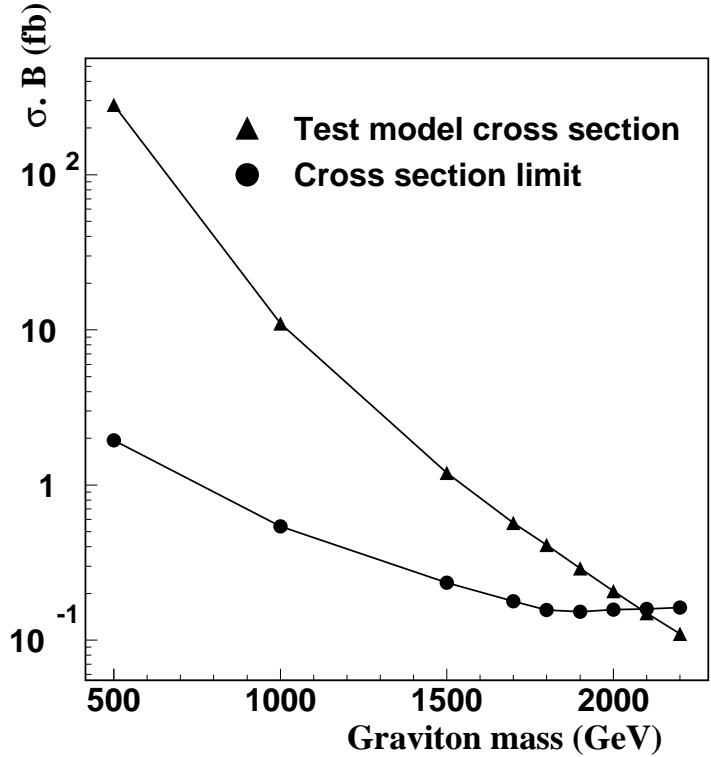
\caption{The cross-section times branching
ratio, $\sigma\cdot B$, for $G\rightarrow e^+e^-$ in the
test model and the smallest detectable cross-section times
branching ratio, 
$(\sigma\cdot B)^{min}$}
\label{sigma}
}

\section{Angular distributions}
In order to demonstrate that the observed resonance is a graviton
and not due to a spin-1 $Z^\prime$, or similar exotic object, it is
necessary to show that it is produced by a spin-2 intermediate
state. 
%The angular distributions expected for $qq\bar\rightarrow G
%\rightarrow e^+e^-$ and $gg\rightarrow G \rightarrow e^+e^-$ are $
%1-5/2 sin^2\theta^* + 2sin^4\theta^*$  and
%$1-cos^4\theta^* $ respectively, where $\theta^* $ is the
%polar angle of the outgoing electron in the rest frame of the
%graviton. 
%The corresponding distribution for both Standard Model background
%and any new spin-1 intermediate state is of the form $ 1+\alpha
%cos^2\theta^*$, where
%$\alpha=1$ for Standard Model processes. 
%These distributions are
%shown in Figure \ref{angdist}a. 
The strong differences between
the angular distributions shown in Figure \ref{ang_theory} would
allow the spin of the resonance to be determined given sufficient
statistics. Information can also be gained on the relative
production rate from the $gg$ and $q\bar q$ processes.

Figure \ref{angdist} shows the angular distribution expected in the test
model for a graviton resonance mass of 1.5 TeV, after passing the events
through the detector simulation and analysis cuts, using the statistics
for 100 fb$^{-1}$  of data. The contributions
from Standard Model Background, $gg$ production and $q\bar q$ production
of the graviton resonance are shown. In this case, 83\% of the graviton
production is from the
$gg$ process. The distribution expected from a spin-1 resonance is
also shown. The cutoff in the detector acceptance at
$|\eta|=2.5$ removes events at large
$|\cos\theta^*|$. For heavy gravitons, which are produced with little
longitudinal momentum, the effect is relatively sharp in
$\cos\theta^*$, while for lighter gravitons and Drell-Yan processes,
the acceptance loss reaches to lower $|\cos\theta^*|$ values.

\FIGURE{
\hbox{\epsfysize=10cm
\epsffile{ang_experiment_1500.epsi}
}
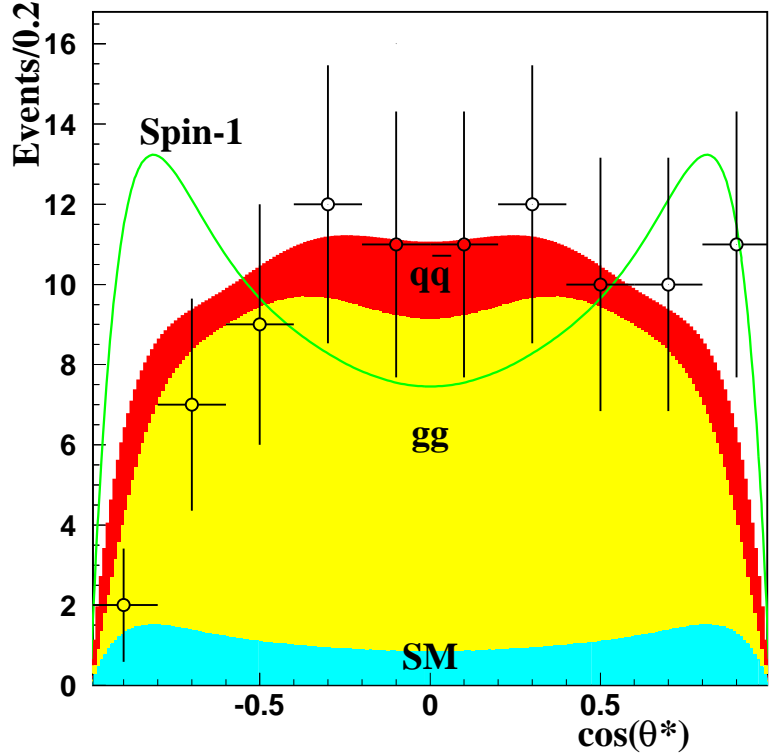
\caption{The angular distribution of data (points with errors) in the test
model for $m_G=1.5$ TeV and 100~fb$^{-1}$ of integrated luminosity. The
stacked histograms show the contributions from the Standard Model (SM), $gg$
production ($gg$) and $q\bar q$ production ($q\bar q$). The curve shows
the distribution expected from a spin-1 resonance.}
\label{angdist}
}

A likelihood function was constructed to quantify the
information in the angular distributions, defined as 
\begin{eqnarray}
L=&x_{q}\cdot f_{q}(\theta^*)\cdot A_{q}(M,\theta^*)/I_{q}(M) 
    +x_{g}\cdot f_{g}(\theta^*)\cdot A_{g}(M,\theta^*)/I_{g}(M)
\nonumber\\
    &+x_{DY}\cdot f_{DY}(\theta^*)\cdot A_{DY}(M,\theta^*)/I_{DY}(M)
%    &+x_{Z'}\cdot f_{Z'}(\theta^*)\cdot
%A_{Z'}(M,\theta^*)/I_{Z'}(M)
\end{eqnarray}
where $x_i$ is the fraction of the events from each contributing
process, $f_i(\theta^*)$ is the angular distribution of the process,
$A_i(M,\theta^*)$ is the acceptance of the detector as a function of
the mass of the electron pair and $\theta^*$, and 
\begin{equation}
I_i(M)=\int_{-1}^{1} f_i(\theta^*)\cdot A_i(M,\theta^*) \ d\cos\theta^*
\end{equation}
$i=q, g, DY$ for the processes $q\bar q\rightarrow G$,
$gg\rightarrow G$, 
%$q\bar q/gg\rightarrow Z'$ 
and
$q\bar q\rightarrow Z/\gamma^*$ respectively. 
%The Z' and Drell-Yan
%processes have identical angular distributions, but are included
%separately in order to allow the expected Drell-Yan background to
%be included explicitly when required. 
Only the shape of the
distribution is used in the statistical tests, and the
coefficients $x$ are constrained such that
\begin{equation}
x_{q}+x_{g}+
%x_{Z'}+
x_{DY}=1
\end{equation}

In order to evaluate the discovery reach of the experiment, in
terms of its ability to reveal the spin-2 nature of the resonance,
the following procedure was followed, intended to mimic an
ensemble of possible experimental runs:
\begin{enumerate}
\item $N$ events were generated, containing the expected
number $N_{DY}$ background events, with a Poisson distribution,
and the remainder being graviton resonance decays with the $q\bar q $ and
$gg$ mixture of the test model at that mass point.
\item $-\log L$ was evaluated for two hypotheses:
\begin{enumerate}
\item $x_{q}=x_{g}=0$,  
%$ x_{Z'}+
$x_{DY}=1 $, corresponding to a spin-1
resonance, such as a $Z'$
\item all the coefficients set to the
values in the test model with the expected Standard Model
background fraction. 
\end{enumerate}
\item $N$ was increased by adding graviton resonance events, until the
difference in $-\log L$ between the two hypotheses reached 1.34,
then 1.92 and finally 3.32. At these points the spin-1 hypothesis
can be eliminated with 90, 95 and 99\% confidence respectively. 
\item This procedure was repeated for many sets of events, and the
average value of $N$ needed for discrimination was found for each confidence
level.
\end{enumerate}
This value of $N$ provides a measure of the statistics needed in a
typical experiment in order to determine the spin-2 nature of the
resonance. $N_S^{min}=N-N_{DY}$ is the number of signal events in the
sample at the confidence limit. Table \ref{ang_events} shows results of
this procedure.

\TABLE{
\caption{The minimum numbers of signal events, $N_S^{min}$, needed to
distinguish the spin-2 and spin-1 hypotheses at 90, 95 and 99\% confidence,
and the corresponding cross-section times branching ratio, for a run of
100 fb$^{-1}$.}
\label{ang_events}
\begin{tabular}{r|ccc|ccc} 
\hline
$m_G$ &         & $N_S^{min}$      &         &&$\sigma\cdot B$
(fb)  &
\\   (GeV) & 90\% CL & 95\% CL & 99\% CL & 90\% CL & 95\% CL &
99\% CL                       \\ 
\hline 
 500 & 140 & 141 & 226 	& 1.90 & 1.92 & 3.07   \\ 
1000 &  59 &  70 &  99	 & 0.80 & 0.94 & 1.33   \\
1500 &  43 & 	48 &  70	 & 0.61 & 0.68 & 1.00   \\
1700 &  41 &  48 &  65	 & 0.61 & 0.71 & 0.96   \\
1800 &  29 &  33 &  58	 & 0.43 & 0.49 & 0.87   \\
1900 &  32 &  40 &  64  & 0.49 & 0.61 & 0.98   \\
2000 &  36 &  41 &  69	 & 0.56 & 0.64 & 1.08   \\
2100 &  31 &  45 &  59	 & 0.49 & 0.71 & 0.93   \\
2200 &  29 &  33 &  55	 & 0.45 & 0.52 & 0.86   \\
\hline
\end{tabular}
}
The 90\% confidence limit for $m_G=500$~GeV is raised by
the requirement that the sample contains at least $5\sqrt{N_{DY}}$ signal
events. In this case, less events would be sufficient
to establish the spin of the resonance, but would not meet our criterion
for a discovery. At very high $m_G$, the discrimination becomes more
difficult because of the increased fraction of the resonance production
coming from the $q\bar q$ initial state.

Figure \ref{ang_limits} shows the confidence limit curves
and the corresponding cross-section times branching ratio in our
test model.  Smooth curves have been drawn through the points since some
statistical fluctuations remain at high mass. These results are model
independent, as long as the assumption that the coupling of the graviton
resonance is univeral holds true. This assumption affects the fraction of
graviton resonance production from
$gg$ and
$q\bar q$ initial states, and hence the shape of the angular distribution. The
cross-section times branching ratio of the test model is shown for comparison.
In this model, the spin of the resonance could be determined with 90\%
confidence for graviton resonance masses as high as 1720 GeV.

\FIGURE{
\hbox{\epsfysize=10cm
\epsfbox{angle_smoothing.epsi}
}
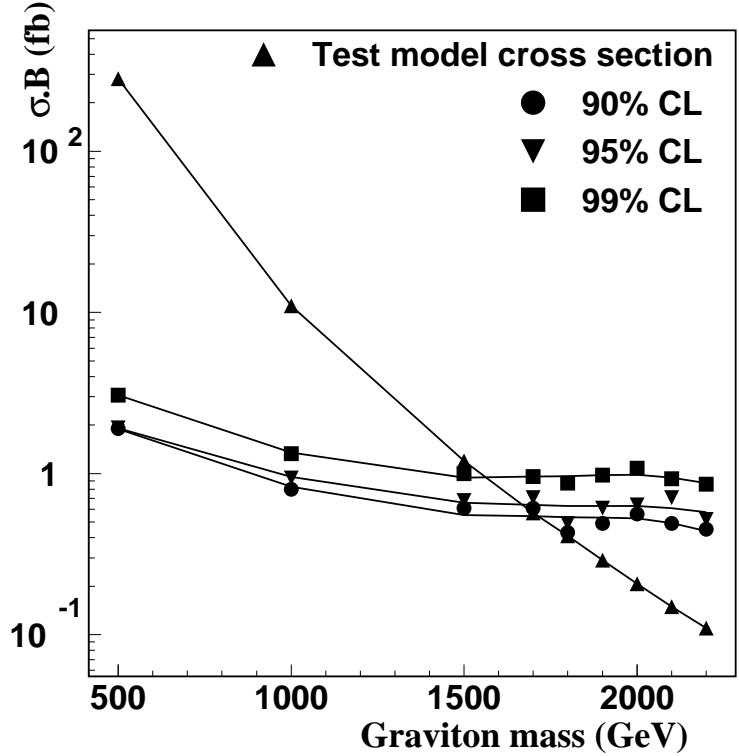
\caption{The smallest cross-section times branching ratio for
which the spin-2 hypothesis is favoured over the spin-1 hypothesis
at 90, 95 and 99\% confidence. The cross-section times branching
ratio of the test model is also shown.}
\label{ang_limits}
}

\section{Conclusions}

The discovery limits for the detection
of the decay mode
$G\rightarrow e^+e^-$, derived above, show that the resonance can
be detected up to a graviton resonance mass of 2080 GeV in our test model.
The limits are model independent as long as the graviton couplings
are universal and give rise to narrow resonances, with widths less
than the experimental resolution. The angular distribution of the
lepton pair can be used to determine the spin of the state. In our
test model, the angular distribution favours a spin-2
hypothesis over a spin-1 hypothesis at 90\% confidence for
graviton masses up to 1720 GeV. 
% from Ben
Once the spin of the resonance has been determined, it will be
useful to check the universality of its couplings~\cite{future}. 
Muon and jet
production
cross-sections in the resonance region as well as 
decays into photons, massive vector bosons, higgs bosons and top
quarks could be used to check this
universality. The bosonic final states would provide a further test of
the spin, since their angular distributions will be different from that of
the electrons considered above. They would also serve to rule out other
models of high mass resonances.  For example, the $Z^\prime$ will not couple to
a two photon final state.

\acknowledgments
We would like to thank J.~March-Russell for an interesting discussion
and I.~Vernon for helpful discussions on different scenarios. MAP
would like to thank C.G.~Lester for his technical help. This work was funded
by the U.K. Particle Physics and Astronomy Research Council.

\end{document}